\documentclass[nato]{crckbked}

\usepackage{graphicx}
\usepackage{klucite}

\makeindex
\newcommand{\be}{\begin{equation}}
\newcommand{\ee}{\end{equation}}
\newcommand{\bea}{\begin{eqnarray}}
\newcommand{\eea}{\end{eqnarray}}

\begin{document}
\numreferences

\begin{article}

\begin{opening}
\title{Competition and coexistence of magnetic and quadrupolar ordering}

% Please give the authors names and email, and put all the authors 
% from the same institution together as  in this example
\author{P. \surname{Fazekas}}%\email{pf@szfki.hu}}
\author{A. \surname{Kiss}}%\email{amk@szfki.hu}}
\institute{Research Institute for Solid State Physics and Optics, \\ 
P.O.B. 49, H-1525 Budapest 114, Hungary}

% Authors, as they appear in the table of contents
\runningauthor{P. Fazekas, A. Kiss}

\begin{abstract}
The large number of low-lying states of $d$- and $f$-shells supports a 
variety of order parameters. The effective dimensionality of the local 
Hilbert space depends on the strength, and kind, of intersite interactions. 
This gives rise to complicated phase diagrams, and an enhanced role of
frustration and fluctuation effects. The general principles are illustrated 
on the example of the effect of a magnetic field on quadrupolar phase 
transitions in some Pr-based skutterudite compounds.   
\end{abstract}
\end{opening}

\section{Introduction}

Transition metal and rare earth compounds show a rich variety of collective
behavior: various kinds of ordered phases as well as strongly fluctuating 
states (spin and orbital liquids). The basic reason is that $d$- and
$f$-shells have a relatively large number of low-energy states. Crystal field
splitting usually reduces this number (the dimensionality ${\cal D}$ 
of the local Hilbert space) considerably below the free ion value, but
complicated physics can arise even from  ${\cal D}=3$ or 4. 

Let us briefly consider some examples. ${\cal D}=3$ is,
in one interpretation, the case of $S=1$ spin models which turned out to have
unanticipated phases like the spin nematics \cite{batista}. A different
realization is offered by interacting $f$-electron models based on the
low-lying quasi-triplet of Pr ions in PrBa$_2$Cu$_3$O$_{7-\delta}$ where the
nature of Pr ordering is still an open issue \cite{prbacuo,amk}. A literal
realization of ${\cal D}=4$ is offered by the $\Gamma_8$ ground state of Ce
ions in CeB$_6$ which has a rich phase diagram \cite{ceb6}. Alternatively, we
may think of the fourfold quasidegeneracy arising from the combination of
twofold spin degeneracy with twofold orbital degeneracy, which is the simplest
model of $d$-electrons which is capable of supporting either spin or orbital
order, or a combination of both. Twofold orbital degeneracy may occur in
cubic, tetragonal, and hexagonal environments, and there are quite different
versions of the $e_g$-model. In the cubic $e_g$ model, orbital momentum is
completely quenched, but orbital order may still break time reversal
invariance in the octupolar phase \cite{octupole}. In contrast, 
the trigonal $e_g$ states
sustain permanent orbital momentum along the threefold axis; such a model
should be relevant for understanding the complex behavior of BaVS$_3$ 
\cite{bavs1,penc1}. The possibilities of ${\cal D}>4$ models are largely 
unexplored, but we should mention the ${\cal D}=6$ $t_{2g}$-models of
LaTiO$_3$. There is an intricate relationship between orbital ordering and
spin ferromagnetism which would be remarkably difficult to explain without due
consideration of the orbital degrees of freedom \cite{FP}. 

As implied above, the definition of ${\cal D}$ is not straightforward: it is
usually higher than the degeneracy of the ground state level (which would
often be small because of small low-symmetry components of the crystal field),
but it is not so high as the free-ion value. In CeB$_6$, the fourfold
degenerate $\Gamma_8$ level is well separated from the higher-lying $\Gamma_7$
which still arises from the Hund's rule ground state; in 
PrBa$_2$Cu$_3$O$_{7-\delta}$, a low-lying doublet and a singlet 
can be lumped together to give a quasi-triplet which would have $\Gamma_5$
character if the symmetry were cubic; but it is, in fact, only tetragonal. In
any case, the relevant dimensionality ${\cal D}=3$ is much smaller than 9
which would be the  Hund's rule value. It depends primarily on the strength 
of intersite interaction, which splittings should be considered small.

The highest symmetry of the the ${\cal D}=3$ models would be SU(3), etc. 
Clearly, the exact realization of a high-symmetry model is more than 
improbable, and if it were really required, we should forget about
it. However, there are indications that the domain of influence of such a
seemingly artificially high symmetry point in parameter space extends over a
substantial portion of the phase diagram \cite{penc1}. 
It stands to reason that SU(${\cal D}$) models (which have symmetries 
connecting spin and orbital axes in Hilbert space) are more quantum 
fluctuating than the pure SU(2) spin models: there are more transverse 
directions to fluctuate to. For instance, the SU(4) model on the triangular
lattice has a plaquette resonating ground state (the SU(4) version of the
resonating valence bond idea \cite{RVB}), and the influence of this
spin--orbital liquid state may extend to physically relevant regions of the 
parameter space \cite{penc2}.

Limitation of space forbids us to present more than one concrete example of
spin--orbital models. We consider Pr-filled skutterudites, in particular 
PrFe$_4$P$_{12}$ which we model as a ${\cal D}=4$ system. 

\section{The case of PrFe$_4$P$_{12}$}

\index{quadrupolar order}\index{ PrFe$_4$P$_{12}$}
Pr-filled skutterudites \index{skutterudites} 
show varied behavior: PrRu$_4$P$_{12}$ has a metal--insula\-tor transition 
\cite{sekine},  PrOs$_4$Sb$_{12}$ is thought to be an exotic 
superconductor \cite{osmium}, while PrFe$_4$P$_{12}$ \index{PrFe$_4$P$_{12}$} 
remains a normal metal in the entire temperature range
studied so far. Our interest lies in PrFe$_4$P$_{12}$ which in a certain 
parameter range can be characterized as a heavy fermion system with 
exceptionally high electronic specific heat \cite{Aoki02}. PrFe$_4$P$_{12}$ 
has a phase transition at $T_{\rm tr}\approx 6.5{\rm K}$ to an ordered phase 
which had first been thought to be antiferromagnetic, but mounting evidence
indicates that it is, in fact, antiferroquadrupolar (AFQ)
\cite{Aoki02,iwasa,curnoe}. \index{antiferroquadrupolar order}
\index{quadrupolar order}

Our purpose is to model the AFQ transition by a crystal field model of the
$4f$ electrons. We assume
that Pr is trivalent ($4f^2$). This can not be literally true, since the
driving force of the formation of a heavy band is presumably the admixture of
other valence states. However, high-field studies show that the heavy 
fermion state competes with AFQ ordering \cite{Aoki02}, so a localized 
$f$-shell description should be acceptable within, or adjacent to, 
the AFQ phase. Besides, at low fields $H$ and high temperatures $T$, 
thermal dehybridization acts to obviate the need to consider interband
coherence effects.

\subsection{The crystal field model}

The $J=4$ manifold of Pr$^{3+}$ is split by the approximately cubic crsytal
field into the $\Gamma_{1}$ singlet, the $\Gamma_{3}$ doublet, and the 
$\Gamma_{4}$ and $\Gamma_{5}$ triplets\footnote{The symmetry group is really 
not $O_h$, but the tetrahedral $T_h$. We nevertheless 
use the cubic classification, which is an approximation at zero field, but
when $H\ne 0$, the symmetry will be in any case substantially lowered.}. 
Group theory does not tell us the sequence of the states, but fitting the
measurements narrows the choice. Analyzing the anisotropy of the magnetization
curves, it was concluded that the likely possibilities are: a  $\Gamma_{1}$
ground state and a low-lying  exited state $\Gamma_{4}$ 
(the $\Gamma_1$--$\Gamma_4$ scheme); or the $\Gamma_1$--$\Gamma_5$ scheme; 
or the $\Gamma_3$--$\Gamma_4$ scheme \cite{Aoki02}. Similar  schemes were
suggested for PrOs$_4$Sb$_{12}$ \cite{osmium,Aokib}.

The assumption that the low-$T$ ordered state is AFQ, seems to speak in favour
of the $\Gamma_3$--$\Gamma_4$ scheme, since then the ionic ground state 
$\Gamma_3$ posesses a (permanent) quadrupolar moment. It was also pointed 
out that the choice of the $\Gamma_3$ ground state is consistent with a
symmetry analysis of the strucrural distortion accompanying the AFQ ordering
\cite{curnoe}. This latter argument relies only on the assumption of the 
$\Gamma_3$ ground state, and does not consider the effects of the low-lying
excited state. Here we show that the assumption of the $\Gamma_1$--$\Gamma_4$
scheme is also capable to account for most of the observed static 
properties of PrFe$_4$P$_{12}$.  

We now discuss the consequences of assuming a $\Gamma_1$--$\Gamma_4$ level 
scheme. Since the singlet ground state  
\be
|\Gamma_1\rangle = \sqrt{\frac{5}{24}}\left( |4\rangle + |-4\rangle \right) 
+ \sqrt{\frac{7}{12}} |0\rangle
\label{eq:g1g4a}\ee
does not carry any kind of moment, the ordered quadrupolar moment
has to be induced by intersite interactions, assuming that the local Hilbert
space contains also the triplet

\begin{eqnarray}
|\Gamma_{4}^+\rangle & = & \frac{1}{4} \left\{ |3\rangle + |-3\rangle +\sqrt{7} \left( |1\rangle + 
|-1\rangle \right) \right\}
\label{eq:g1g4b}\\[2mm]
 |\Gamma_{4}^0\rangle  & = & 
\frac{1}{\sqrt{2}}\left( |-4\rangle - |4\rangle \right)\label{eq:g1g4c}\\[2mm]
|\Gamma_{4}^-\rangle & = &  \frac{1}{4} \left\{ |3\rangle - |-3\rangle +\sqrt{7} \left( |-1\rangle - 
|1\rangle \right) \right\}\, , \label{eq:g1g4d}
\end{eqnarray}
where we have chosen the basis of quadrupolar eigenstates. 
Choosing the energy of (\ref{eq:g1g4a}) as the zero, the states 
(\ref{eq:g1g4b})--(\ref{eq:g1g4d}) lie at the level $\Delta$.

The possible moments in the four-dimensional local Hilbert space
spanned by (\ref{eq:g1g4a})--(\ref{eq:g1g4d}) are given by the decomposition 
\be
(\Gamma_1\oplus\Gamma_4)\otimes(\Gamma_1\oplus\Gamma_4) = 
2\Gamma_1+\Gamma_3+3\Gamma_4+\Gamma_5\, .
\label{eq:decomp1}
\ee
Evidently, the system could support either dipolar ($\Gamma_4$), or either of 
two kinds of quadrupolar ($\Gamma_3$ or $\Gamma_5$) order\footnote{This is a
  classification of order parameters which can be defined purely
  locally. $Q\ne 0$ order needs further discussion.}. The quadrupolar
order parameters are the same that appear in the decomposition of a purely
$\Gamma_4$ system  
\be
\Gamma_4\otimes\Gamma_4 = \Gamma_1+\Gamma_3+\Gamma_4+\Gamma_5
\label{eq:decomp2}\, ,
\ee
i.e., they are not sustained by inter-level matrix elements\footnote{In
  contrast, matrix elements between $\Gamma_1$ and $\Gamma_4$ would bring
  extra possibilities of dipolar ordering. This does not seem to be relevant
  for PrFe$_4$P$_{12}$.}. -- It may be of some interest to mention that the 
$\Gamma_1$--$\Gamma_4$ scheme does not offer the possibility of octupolar 
order (but the $\Gamma_3$--$\Gamma_4$ scheme would).

\subsection{The effect of external magnetic field}

Our decomposition (\ref{eq:decomp1}) allows to seek dipolar and/or quadrupolar
ordering in the system. Experiments give the clue that we should, in fact,
look for (antiferro)quadrupolar order. We may rather arbitrarily assume that
it is of the $\Gamma_3$ kind\footnote{Assuming ${\cal O}_{xy}$-type $\Gamma_5$
  ordering would give similar results.}, i.e., the possible order parameters
are ${\cal O}_2^0=3J_z^2-J(J+1)$ and ${\cal O}_2^2=J_x^2-J_y^2$. 
Furthermore, since the total energy expression for a pair of sites has only
tetragonal (as opposed to cubic) symmetry, we need not assume that the 
${\cal O}_2^0$ and ${\cal O}_2^2$ couplings would be equal, and we may seek, 
say, ${\cal O}_2^2$-type order. Using a mean field decoupling, our problem
would be rather similar to a four-state Blume--Emery--Griffiths model.

It is a well-known feature of quadrupolar ordering that its phenomenology
closely imitates that of antiferromagnetic transitions, though the underlying
order parameter is non-magnetic. The phase diagram in the $H$--$T$ plane ($H$:
magnetic field) was mapped in \cite{Aoki02}. The salient features are the
following: A sufficiently strong field applied in any direction will suppress
AFQ ordering completely. On the phase boundary, a low-field regime of
continuous transitions is separated by a tricritical point 
($H^*\approx 2{\rm Tesla}$, $T^*\approx 4{\rm K}$) from the high-field regime
of first-order transitions. This change in the character of the phase
transition is shown in the field dependence of the specific heat. 
The nature of the magnetization curve changes \index{metamagnetic transition} 
drastically at $T^*$. For $T<T^*$, there is a steplike metamagnetic transition
corresponding to the first-order transition from the low-$T$ ordered phase to the
disordered phase. For $T^*<T<T_{\rm tr}(H=0)\approx 6.5{\rm K}$, there is a
kind of a smooth metamagnetic transition, where the system crosses the
second-order part of the phase boundary. For $T>T_{\rm tr}(H=0)$, the
magnetization curve is completely smooth. We will show that a mean field
treatment of the AFQ transition in the $\Gamma_1$--$\Gamma_4$ scheme accounts
for these observations quite well.
  
In the absence of an external magnetic field, $\Gamma_3$ and $\Gamma_4$ type
order parameters (i.e., quadrupolar moment and magnetization) are decoupled
because the former is invariant under time reversal, while the latter changes
sign. Switching on the magnetic field breaks time reversal invariance, allowing that
quadrupolar moment and magnetization get mixed. We can also say that magnetic
field, though it couples directly to the angular momentum ${\vec J}$, may also
induce quadrupolar moment.

This is best illustrated by looking at the matrix which contains the matrix
elements of the crystal field, the Zeeman energy for a field in the
$x$-direction $-h_xJ^x$, and also a term containing 
the quadrupolar moment $\lambda{\cal O}_2^2$, within the basis 
(\ref{eq:g1g4a})--(\ref{eq:g1g4d}) 
\be
{\cal M}(\lambda,h_x,\Delta) = \left( \begin{array}{cccc}
           0 & -2\sqrt{5/3}h_x & 0 & 0 \\
           -2\sqrt{5/3}h_x & \Delta+7\lambda & 0 & 0 \\
           0 & 0 & \Delta & -h_x/2         \\
           0 & 0 & -h_x/2 & \Delta-7\lambda         
           \end{array}
       \right)
\label{eq:matrix}
\ee
The field couples the singlet ground state $|\Gamma_1\rangle$ to the 
${\cal O}_2^2$-moment bearing excited state $|\Gamma_4^+\rangle$. Therefore in
the presence of a magnetic field, uniform quadrupolar moment is no longer
"spontaneous". If at $H=0$, we had to do with a transition to a
ferroquadrupolar state, it would be smeared out in $H\ne 0$, and we no longer
had a phase boundary to speak about\footnote{Purely in symmetry terms:
  applying a field in one of the cubic (100) directions, the symmetry would be 
  lowered to $C_{4h}$, and the decomposition of $\Gamma_4$ of $O_h$ in
  terms of the irreps of $C_{4h}$ would contain the identity which also
  comes from $\Gamma_1$ ground state.}. However, for  
  {\sl anti}ferroquadrupolar coupling, the appearance of the staggered
  quadrupolar moment is still symmetry breaking, and therefore a sharp phase
  transition remains possible also in an external magnetic field. Therefore,
  if we had no other evidence than that PrFe$_4$P$_{12}$ has sharp phase
  transitions in external magnetic field, and we adopted the
  $\Gamma_1$--$\Gamma_4$ scheme, we would have to conclude that the ordered
  state could not be ferroquadrupolar, but only antiferroquadrupolar.
 
\subsection{Mean field results}

Doing the mean field theory \cite{amk} for AFQ ordering involves diagonalizing matrices
like (\ref{eq:matrix}), and we do not give the details here. 

\begin{figure}
\centerline{\includegraphics[height=6cm,angle=270]{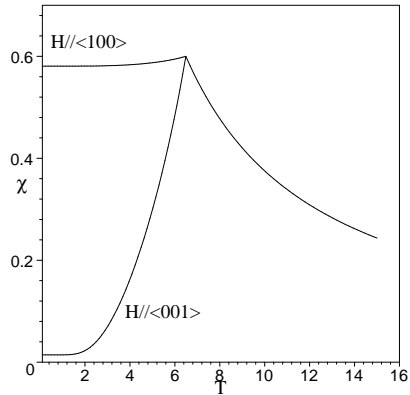}}
\caption{The temperature dependence of the magnetic susceptibility. The onset
  of ${\cal O}_2^2$-type quadrupolar order makes the two $H\parallel z$ and
  $H\parallel x$ behavior inequivalent.}\label{fig:figsusc}
\end{figure}

Fig.~\ref{fig:figsusc} illustrates that an AFQ transition, 
though of non-magnetic
nature,  may give a susceptibility which looks 
very much like what you expect from an
antiferromagnet. The cubic (001) and (100) directions are 
equivalent in the para phase but the appearance of ${\cal O}_2^2$-type AFQ
order makes the $x$-field susceptibility appear as "transverse", while the  
$z$-field susceptibility looks "longitudinal". Of course, instead of 
${\cal O}_2^2=J_x^2-J_y^2$ we might have chosen $J_y^2-J_z^2$ or
$J_z^2-J_x^2$, so in a crystal one would expect an equal mixture of the
corresponding AFQ domains, and the susceptibility suitably averaged. The
experiments may be taken to correspond to this. 

\begin{figure}
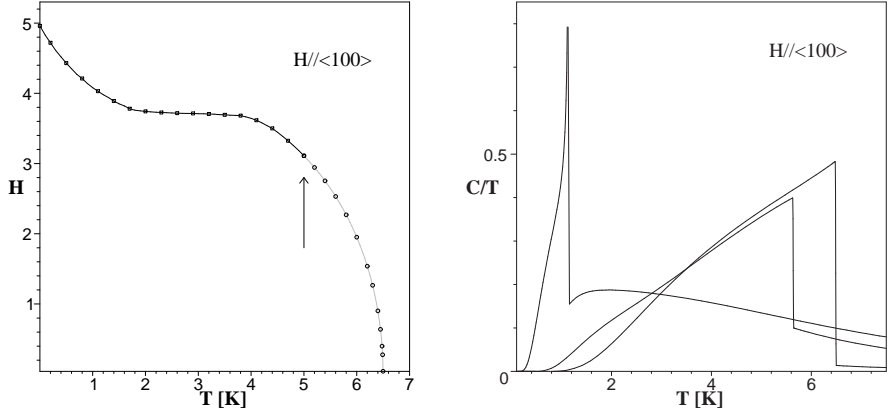

\centerline{\includegraphics[width=6.4cm,angle=270]{fazisd09.ps}
\includegraphics[width=6.4cm,angle=270]{fajho01.ps}}
\caption{{\sl Left}: The boundary of the antiferroquadrupolar phase in the 
$H$--$T$ plane. The curve is drawn in black for first-order transitions, and 
in grey for continuous phase transitions. Arrow indicates the  tricritical
point. {\sl Right}: The $T$-dependence of the temperature coefficient of the 
specific heat for $H=0$, 2.5, and 4Tesla (in order of decreasing transition temperatures).}\label{fig:figphase}
\end{figure}

Fig.~\ref{fig:figphase} (left) gives the phase diagram. The crystal field
splitting $\Delta$ and the quadrupolar coupling $\lambda$ were chosen so as to
get at least a rough numerical agreement with the experimental phase diagram
\cite{Aoki02}. There is still some freedom in the parameters, but we found
that a rather low $\Delta\approx 4-6{\rm K}$ has to be chosen (with $z=8$,
$\lambda\approx 0.08k_{\rm B}$), if we want to get both $T_{\rm tr}(H=0)$ and
the tricritical point right. These estimates are likely to be subject of some
revision when further (especally dipolar) couplings are allowed for. -- In
spite of an overall similarity to the phase diagram based on experiments, we
note that the low-$T$, high-$H$ upcurving part of our present phase boundary
represents a deviation, the reason for which remains to be clarified.

\begin{figure}
\centerline{\hspace{0.5cm}\includegraphics[width=12.0cm]{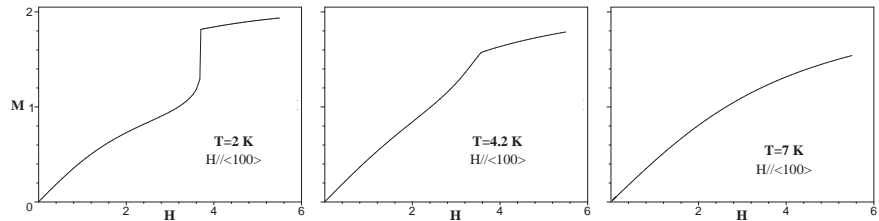}}
\caption{Antiferroquadrupolar order underlies the sharp metamagnetic 
transition at low $T$ (left). With increasing $T$, we pass through second
order transitions (middle) to smooth behavior (right).}\label{fig:figmeta}
\end{figure}

The changeover to a regime of first-order transitions in higher fields is
evident in the field dependence of the specific heat; the curves shown in 
Fig.~\ref{fig:figphase} (right) bear a close resemblance 
to the measured ones. The
same is true of the magnetization curves (Fig.~\ref{fig:figmeta}) 
where we see a
change from the regime of sharp metamagnetic transitions at low temperatures
to continuous phase transitions at intermediate $T$'s, and eventually smooth
behavior in the para phase.  
   
\begin{acknowledgements}
The authors profited much from cooperation with K. Penc, L. Forr\'o,
G. Mih\'aly and I. K\'ezsm\'arki, and from discussions with H. Shiba. 
This work was supported by the grants OTKA T038162, and AKP 2000-123 2,2. 
\end{acknowledgements}

\end{article}

\end{document}